\documentclass[twocolumn,showpacs,preprintnumbers,amsmath,amssymb]{revtex4}
\usepackage{amssymb}     
\usepackage{graphicx}    
\usepackage{dcolumn}
\usepackage{bm}

\begin{document}

\title{Gravitational waves from galaxy encounters}

\author{Vicent Quilis$^{1}$, A. C\'esar Gonz\'alez-Garc\'{\i}a$^{2,3}$,
        Diego S\'aez$^{1}$ and Jos\'e A. Font$^{1}$}
        
 \affiliation{  $^{1}$Departament   d'Astronomia  i  Astrof\'{\i}sica,
    Universitat  de   Val\`encia,  Dr.~Moliner  50,   46100  Burjassot
    (Val\`encia),  Spain\\   $^{2}$Instituto  de  Astrof\'{\i}sica  de
    Canarias, V\'{\i}a L\'actea s/n, La Laguna, 38200, Spain \\
    $^{3}$Departamento de F\'{\i}sica Te\'orica, C-XI, 
    Universidad Aut\'onoma de Madrid,  Madrid, 28049, Spain}

\date{\today}

\begin{abstract}
We  discuss  the  emission  of  gravitational  radiation  produced  in
encounters of  dark matter galactic halos.   To this aim  we perform a
number of  numerical simulations of typical  galaxy mergers, computing
the associated gravitational radiation waveforms as well as the energy
released  in  the  processes.   Our  simulations  yield  dimensionless
gravitational  wave   amplitudes  of  the  order   of  $10^{-13}$  and
gravitational wave frequencies of the order of $10^{-16}$ Hz, when the
galaxies are located at a distance  of 10 Mpc. These values are of the
same order as those  arising in the gravitational radiation originated
by strong variations of the gravitational field in the early Universe,
and therefore, such gravitational waves cannot be directly observed by
ground-based  detectors.  We  discuss the  feasibility of  an indirect
detection by means of the  B-mode polarization of the Cosmic Microwave
Background  (CMB) induced  by such  waves. Our  results show  that the
gravitational  waves from  encounters  of dark  matter galactic  halos
leave much too small an imprint on the CMB polarization to be actually
observed with ongoing and future missions.
\end{abstract}


\maketitle

\section{Introduction}

Galaxy encounters and interactions have well-established observational
evidence.   Supermassive  black holes  (SBHs)  in  the  mass range  of
$10^6-10^9$ M$_\odot$  are firmly  believed to exist  at the  cores of
AGNs and quasars.  During such interactions gravitational radiation is
going  to  be  emitted.  As  a  result,  a  stochastic  background  of
gravitational waves  (GWs) from SBH encounters  and coalescence sweeps
the  cosmos, with wave  frequencies of  the order  of $10^{-6}$  Hz or
below.  The gravitational waves produced by the merger of binary SBHs,
in terms of both the amplitude  of the signal as well as the frequency
range,  are amenable to  detection by  the Laser  Interferometer Space
Antenna (LISA), becoming one of the most promising anticipated targets
with an optimist signal-to-noise ratio of about 1000~\cite{holz04}.

Binary  black  hole evolution  and  merger  has received  considerable
numerical attention  in recent years through  N-body simulations (see,
e.g.~\cite{merritt05,merritt06}  and  references  therein).   Examples
include  galaxy  mergers  in  which the  pre-merger  galaxies  contain
power-law nuclear  cusps and massive particles  representing the SBHs.
Previous work  on the GWs  emitted from such galaxy  interactions have
only considered  the radiation produced during the  coalescence of the
central SBHs~\cite{enoki04,sesana05}. Thus  far, however, no attention
has yet been paid to the gravitational radiation emitted from the bulk
of  matter of  the interacting  galaxies, i.e.~neglecting  the central
SBHs, notwithstanding the fact that  the mass of the galaxy which lies
outside the  black hole is  typically some orders of  magnitude larger
than the  SBH mass itself.  The  underlying reason for  the absence of
investigations in such aspect of galaxy encounters may have to do with
the fact that  the size of the coalescing  objects involved (the whole
galaxies)  is  not compact  enough  to  yield  a direct  observational
measure in the frequency range of the upcoming interferometer LISA.

On the other hand the dominant component accounting for the total mass
of the galaxy is the  dark matter part. Hence, an interesting question
to   ask  oneself  is   whether  a   hypothetical  detection   of  the
gravitational  radiation emitted from  galaxy encounters,  modelled as
interactions of  (collisionless) dark matter halos lacking  the SBH at
the centre, would provide an  unambiguous signature of the presence of
such  elusive matter. If  that were  the case  the detection  of those
GWs would then provide  a smoking-gun signature of yet another
traditionally  evasive concept in  gravitational physics,  namely dark
matter.

Our aim in  this work is to present  and discuss numerical simulations
of a reduced  but representative sample of such  galaxy encounters and
to    compute    the    associated   gravitational    waveforms    and
luminosities.  Building  on  these  results we  conjecture  about  the
detectability of such signals by electromagnetic means, namely through
their imprint  on the polarisation of  the Cosmic Microwave Background 
(CMB hereafter).

The  non-linear  evolution  of  cosmological scalar  perturbations  as
sources    of    the    CMB    polarisation   has    been    discussed
by~\cite{mollerach04}     using     perturbation     theory.      More
recently~\cite{carbone06} have reported a related study which focus on
the cosmological  stochastic GW background produced  by idealised cold
dark  matter   halos  via  power  transfer  from   scalar  and  vector
perturbations to  tensor metric  modes, during the  strongly nonlinear
stage of their evolution. Since  the nonlinear evolution of such halos
occurs  on  a  cosmological  timescale, the  associated  gravitational
radiation may  be relevant at  frequencies comparable to those  of the
primordial GW which  affect the CMB photons and  produce secondary CMB
anisotropy and polarisation. While the idea behind our work shares the
views of~\cite{carbone06}, the GWs we analyse arise from galactic-like
halos  of  dark  matter,  which  are  substantially  more  asymmetric.
Moreover,  our  study focus  on  the  particular  episode of  galactic
mergers,  where the geometrical  features of  the process  produce GWs
with  distinctive (burst-like)  waveforms and  larger  amplitudes than
those considered  by~\cite{carbone06} and therefore,  their effects on
the CMB polarisation are likely to have greater significance.

Our  study  is  further   motivated  by  the  prospects  of  potential
experimental measurability.   There are a number of  missions aimed to
detect signatures of  GWs (with wavelengths comparable to  the size of
the  Universe) produced  by quantum  fluctuations of  spacetime during
inflation,  by measuring  the weak  imprint they  leave on  the B-mode
polarisation    of    the    CMB.     There    have    already    been
claims~\cite{baskaran06} suggesting  that the polarisation  induced on
the CMB  by relic gravitational waves  at large scales  can be already
measured on the available data  from WMAP, a situation which is likely
to  be improved by  the Planck  mission. According  to~\cite{hu02} the
B-mode  polarization  spectrum of  the  CMB  due  to both,  primordial
(inflation) gravitational waves  and {\it also} gravitational lensing,
is well within the range of detectability of the Planck satellite.  In
addition,  several experiments  have been  proposed for  measuring the
polarisation (E and  B modes) of the CMB  with higher sensitivity than
WMAP (see e.g.~\cite{bowden04}). Specially suited for the detection of
B-mode  polarisation induced by  relic GWs  are the  upcoming projects
CLOVER~\cite{maffei05} and BICEP~\cite{keating03}.  Furthermore, among
the  missions included in  NASA's {\it  Beyond Einstein}  program (see
{\tt universe.nasa.gov}) the Cosmic  Inflation Probe also appears well
suited  to investigate  the  ultra-low frequency  GWs  from the  early
Universe.

The simulations  reported in this  paper show that the  frequencies of
the gravitational  waves produced  by galaxy mergers  ($\sim 10^{-16}$
Hz)  match  the frequency  sensitivity  window  of the  aforementioned
future missions (namely the  Cosmic Inflation Probe). The amplitude of
these GW signals is likely to be of comparable order than those coming
from primordial waves originated  during the early Universe.  However,
we estimate that the CMB polarization they produce is weaker than that
due  to primordial  waves.   While the  imprint  of the  gravitational
radiation from galaxy mergers  on the CMB polarization is inaccessible
to detection, the gravitational lensing effect associated with gravity
perturbations   produced  in  our   simulated  galaxy   mergers  could
marginally  lie within  the  range  of detectability  of  a number  of
planned  missions designed  to measure  such polarization.  This issue
would be discussed elsewhere.

The paper is organized  as follows: In Section~\ref{model} we describe
the  galaxy model,  the setup  of the  numerical simulations,  and the
numerical  code  used.  In  Section~\ref{gw}  the  basic concepts  and
equations   used  to  extract   gravitational  waveforms   within  the
(Newtonian)  slow-motion approximation are  described. The  results of
our  investigation are  presented in  Section~\ref{results}.   Finally, 
Section~\ref{summary} outlines the summary of this research.

\section{Galaxy model and numerical approach}
\label{model}

Galaxy  mergers   are  believed  to  be  a   formation  mechanism  for
present-day elliptical  galaxies~\cite{toomre72}.  Different formation
scenarios have  been claimed to form high  luminosity boxy ellipticals
and  low  luminosity disky  ellipticals.   Boxy  ellipticals could  be
formed     through     dry      mergers     of     bulge     dominated
progenitors~\cite{GG05a,GG05b,khochfar05,burkert05}     while    disky
ellipticals could be formed through unequal mass merging of disks with
a  prominent bulge~\cite{GG05c}  or by  including gas  in  the initial
systems~\cite{naab06}.

For the  present study we  adopt the collisionless approximation  as a
reasonable approach and simulate the merger of two elliptical galaxies
and the merger  of two disks with three  different configurations (see
Table~\ref{tab1}).   The initial conditions  for the  models described
here   are   taken    from   the   samples   studied   by~\cite{GG05a}
and~\cite{GG05c}.  We re-simulate three  key experiments with a larger
number of  particles.  We choose  those experiments that  produced the
largest asymmetries  during the merging  stages for the  remnants with
realistic properties as compared to real-life ellipticals.

To build and  run the simulations we adopt  non-dimensional units with
$G=1$ for  Newton's constant of  gravity.  We use  isotropic spherical
Jaffe models~\cite{jaffe83}  as initial conditions  for the elliptical
galaxies   (see~\cite{GG05a}  for  a   detailed  description   of  the
algorithm). The  projected surface density  of such models  exhibits a
slope that decreases with the radius roughly as $R^{1/4}$, which makes
it a suitable representation  for elliptical galaxies.  Jaffe model is
characterised by  the theoretical half mass radius  $r_{\rm J}$, which
we choose  to be equal to 1.   Because a cut-off radius  is imposed at
$r= 10 \times r_{\rm J}$, the true half-mass radius $r_{1/2}$ is 0.82.

We  follow   ~\cite{Kuijken95}  to  build  the   initial  disk  galaxy
models.  Our initial  models have  a bulge,  a disk,  and a  halo. The
initial  parameters for  these models  are identical  to  models $dbh$
in~\cite{GG05c} but with a larger  number of particles. The model with
a  different mass  is  a scaled  up  version of  the  low mass  system
following a Tully-Fisher like relation

For scaling our  Jaffe models to elliptical galaxies  the set of units
 is:  $[M]  = M_{\rm  J}  =  4\times10^{11}  \; \rm{M_{\odot}}$;  $[L]
 =r_{\rm  J} = 10  \;\rm{kpc}$; $[T]  = 2.4\times10^7  \;\rm{yr}$, and
 finally $[v] =  414\; \rm{km/s}$. A set of  physical units that match
 the {\it dbh} models to the Milky Way are: $[M] = 3.24 \times 10^{11}
 \;   \rm{M_{\odot}}$;   $[L]  =   14.0   \rm   \;   {kpc}$;  $[T]   =
 4.71\times10^7\; \rm{yr}$, with $[v] = 315 \rm \; {km/s}$.

Table~\ref{tab1} reports the orbital parameters of the simulations. We
first consider two equal-mass mergers between elliptical galaxies with
different impact  parameters.  Model  M1 is a  head-on collision  on a
parabolic orbit,  while model  M2 is a  grazing encounter with  a fair
impact parameter,  the orbital angular momentum is  thus larger.  Each
elliptical galaxy  is modelled with $4\times10^5$  particles for model
M1 and with $10^5$ particles for  run M2. Model M3, on the other hand,
is a  merger between  two disk  galaxies with a  mass ratio  3:1. Both
systems  are placed  on penetrating  prograde orbits.  The  orbits are
elliptical ($e=0.7$)  to ensure merging and  avoid large computational
times.  The two disks are placed with their spin slightly coupled with
the orbital angular momentum.  Each disk galaxy simulation consists of
$2.75  \times  10^5$  particles  for  the  halo,  $1.75  \times  10^5$
particles for the disk and $9 \times 10^4$ for the bulge. This makes a
total of  $10^6$ particles for the  merger remnant. For  all models of
our sample, after  merger is completed the remnants  are let to evolve
in isolation for 10 half-mass radius crossing times to ensure that the
inner parts are close to virial equilibrium.

\begin{table}
\caption{Initial configurations  for the  models. (1) Model  name, (2)
type of  merger, elliptical-elliptical  (E+E) or disk-disk  (S+S), (3)
Mass  ratio,  (4)  initial   orientation  of  the  spin,  (5)  initial
separation, (6) orbit ellipticity and (7) impact parameter.}
\label{tab1}
\begin{ruledtabular}
\begin{tabular}{ccccccc}
 {\bf    Mod.}    &{\bf    Type}&{\bf   $    \frac{M_1}{M_2}$}&   {\bf
$(\theta_1,\phi_1)$  $(\theta_2,\phi_2)$} & {\bf  $r_i$} &  {\bf $e$}&
{\bf $D$}\\ (1) & (2) & (3) & (4) & (5) & (6) & (7)\\ \hline $M1$& E+E
& 1:1 & (0,0);(0,0) & 40 & 1 & 0 \\ $M2$& E+E & 1:1 & (0,0);(0,0) & 30
& 1 & 4.9 \\ \hline $M3$ & S+S & 3:1 & (10,-10);(70,30) & 12.4 & 0.7 &
3.72 \\
\end{tabular}
\end{ruledtabular}
\end{table}

We  have  performed  the  simulations  with the  parallel  version  of
GADGET~\cite{springel01}   using   16  CPU's   on   the  IAC   Beowulf
Cluster. Each run takes of the order of $5 \times 10^5$ seconds of CPU
time.   Independent softening is  applied for  each particle  type.  A
tolerance  parameter  of  $\theta  =  0.8$, quadrupole  terms,  and  a
variable time step have been used. Energy conservation is satisfactory
as variations are kept below the $0.5 \%$ level.

\section{Gravitational radiation}
\label{gw}

The  study presented  in this  paper is  focused in  galaxy encounters
taking place, in a flat Universe,  at distances ($D< 100$ \, Mpc) much
smaller than the horizon scale.   The galactic systems we consider are
far from  being relativistic in both senses,  special relativity ($v/c
\le   10^{-3}$),  and   general  relativity   ($r_g/R   \le  10^{-4}$,
$r_g=2GM/c^2$ being the Schwarzschild radius of the object).

Under these  conditions the  gravitational radiation emitted  from the
encounters  of our  galaxy halos  can  be described  by the  so-called
slow-motion formalism, where the spacetime metric can be linearised in
the usual way  ($g_{\mu \nu} = \eta_{\mu \nu} +  h_{\mu \nu}$) and the
transverse  traceless (TT) gauge  can be  used everywhere  outside the
system~\cite{MTW}.   The spatial  components of  $h_{\mu \nu}^{^{TT}}$
are
\begin{equation}
h_{ij}^{^{TT}}=\frac{G}{c^4}\frac{2}{D}\ddot{I}_{ij}
\left(t-\frac{D}{c}\right)
\label{hij}
\end{equation}
where $I_{ij}$ are the components of the traceless inertial tensor.

The collisionless  particle contribution to  $I_{ij}$ is given  by the
following summation extended over all matter particles,
\begin{equation}
I_{ij}=  \sum_{n_p}  m_p  \left(x_i   x_j  -  {1\over  3}  \delta_{ij}
x^2\right),
\end{equation}
$n_p$ being the number of  particles, $m_p$ the mass of each particle,
and $\delta_{ij}$ Kronecker's delta.
 
Outside  the  galaxies,  the   relative  motion  of  two  neighbouring
particles  A and  B  moving with  the  cosmological background  (ideal
detector) is  fully described by the  quantities $h^{^{TT}}_{ij}$.  In
the TT gauge  there is a system of coordinates attached  to A in which
the    coordinate     variations    of    the     particle    B    are
$X^{i}_{_{B}}(t)-X^{i}_{_{B0}}=\frac{1}{2}X^{j}_{_{B0}}
[h^{^{TT}}_{ij}(t)]_{_{A}}$,  where  $X^{i}_{_{B0}}$  stands  for  the
initial   coordinates  of   the  particle   $B$  and   the  quantities
$h^{^{TT}}_{ij}$ are  calculated at point  $A$. From this  formula, it
follows that  oscillations in the $h^{^{TT}}_{ij}$  quantities lead to
oscillations  in the  relative  position  of particles  A  and B  with
related frequencies and amplitudes.

In  the  slow  motion  approximation,  the  gravitational  luminosity,
$L_{GW}$, is given by the well known formula
\begin{equation}
\label{lum}
L_{_{GW}}={G\over{5c^{5}}}\sum_{ij}
{\stackrel{\displaystyle{...}}{I}}_{ij}
{\stackrel{\displaystyle{...}}{I}}_{ij} \ .
\end{equation}

\begin{figure*}                                 
\centering \scalebox{0.75}{\includegraphics{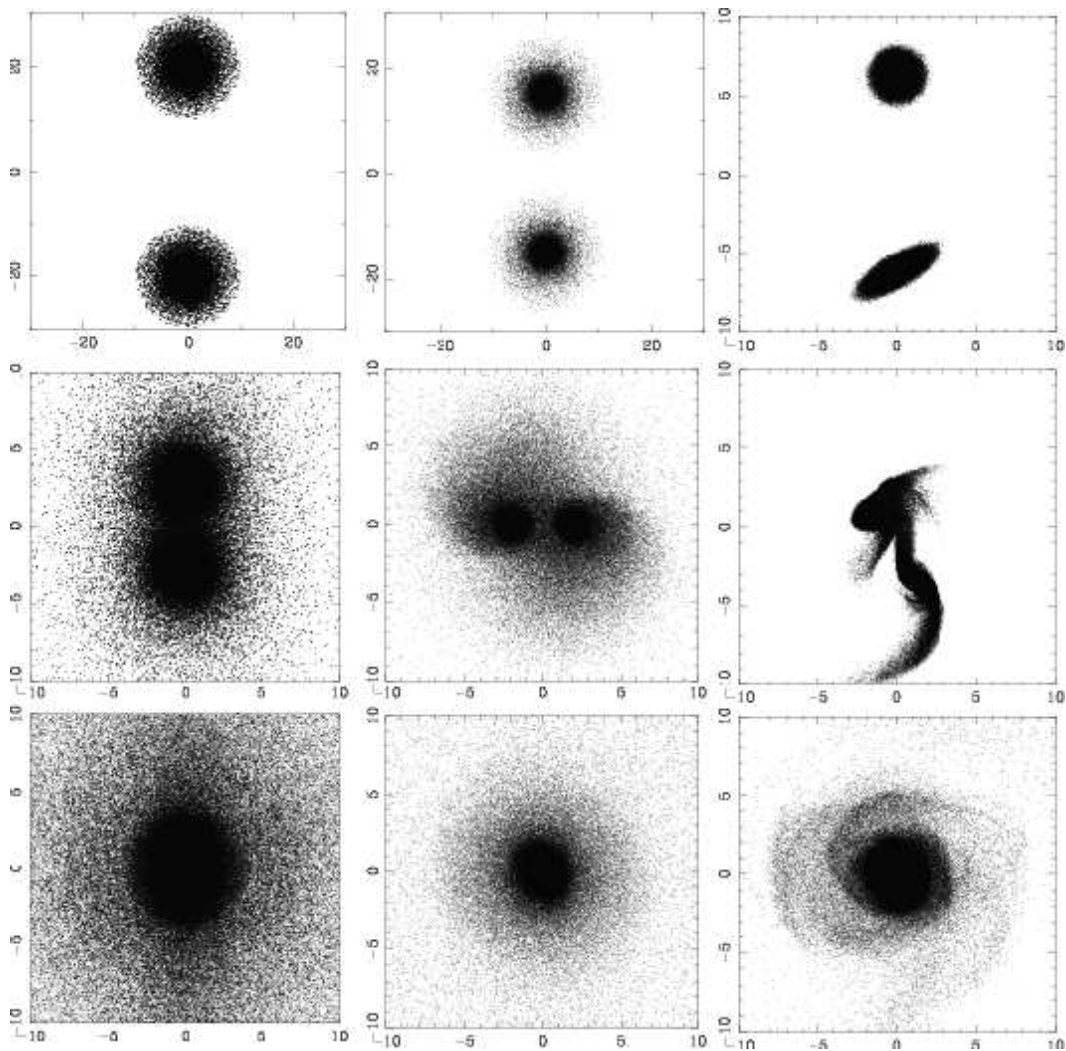}}
\caption{Dynamical evolution  of the three merger  simulations as seen
from a point  of view parallel to the  orbital angular momentum.  Left
column:  M1, middle  column: M2,  right column:  M3.  Top  row initial
conditions, middle row right  after the first pass through pericenter,
bottom row at the end of the simulation.\label{fig1}}
\end{figure*}

\begin{figure}
\includegraphics[width=77mm]{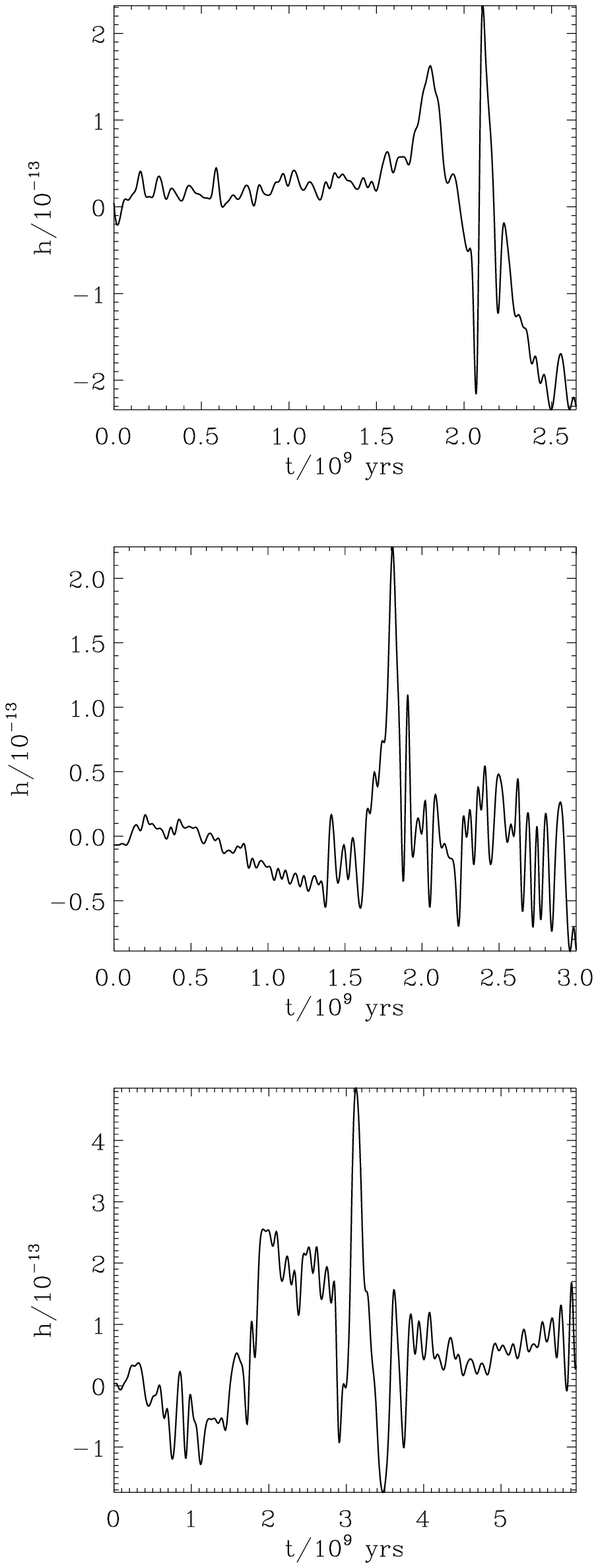}
\caption{Gravitational  waveform for  the three  models: M1  (top), M2
(middle),  and  M3  (bottom).   Only  the ``+"  polarisation  mode  is
shown. Galaxies are located at $D=10$ Mpc from the observer.}
\label{fig2}
\end{figure}

A  direct   calculation  of   $L_{GW}$  based  on   Eq.(\ref{lum})  is
problematic~\cite{finn90} as  a result of the noise  introduced in the
numerical  computation of third-order  time derivatives.   However, as
several          authors         have          suggested         (see,
e.g.~\cite{moenchmeyer91,dimmelmeier02}),    the   second-order   time
derivatives involved in $\ddot{I}_{ij}$ can be written in terms of the
quantities  $\ddot{x}$  and $\dot{x}$,  which,  in  its  turn, can  be
expressed in terms of related  variables using the equations of motion
of the system.

It is well-known that, in  the case of gravitational waves, the metric
perturbation   $h_{ij}$  only  depends   on  two   independent  linear
polarisation  states. Any electromagnetic  radiation propagating  in a
spacetime perturbed  from Minkowski  by the presence  of gravitational
radiation,  will  suffer  such  polarisation itself.   A  particularly
interesting case is represented by  the CMB, and it is indeed possible
to take advantage of the  polarisation of the CMB to design strategies
to detect gravitational  waves~\cite{seljak97,baskaran06}. The case of
relic,  ultra-low frequency  GWs has  been recently  studied  in great
detail by~\cite{baskaran06}.  The order of magnitude  of the amplitude
and frequency corresponding to  the present-day primordial GW spectrum
is $\sim 10^{-15}-10^{-5}$  and $\sim 10^{-20}-10^{-15}$, respectively
(see Fig.~2 in~\cite{baskaran06}).

We will show in the  following section that the source investigated in
our  work, galaxy encounters,  leads to  values of  the GW  strain and
frequencies which overlap to those arising in strong variations of the
gravitational field in the early Universe.

\section{Results}
\label{results}

\subsection{Dynamics of the encounters}

We begin the discussion of  our results by analysing the morphological
features which  arise in our simulations of  galaxy mergers, deferring
for the next section the implications of such dynamics in the emission
of gravitational radiation.

Figure~\ref{fig1}  displays  three snapshots  of  the three  simulated
merger  encounters.   Model  M1  (Fig.~\ref{fig1} left  column)  is  a
head-on encounter between two equal-mass elliptical galaxies placed on
an  orbit with parabolic  orbital energy.   The two  galaxies approach
each other  and there  is not much  disturbance while they  get closer
prior to the  first pass through the pericenter  (which corresponds to
the middle row of Fig.~\ref{fig1}),  apart from a heating of the outer
parts.  A  large fraction of the  orbital energy is lost  in the first
encounter,  a  number  of  particles  are expelled  from  the  system,
becoming  unbound  and  probably  forming  part of  the  unbound  free
floating    intergalactic   population~\cite{stanghellini06}.    These
particles carry out a fraction  of the orbital energy away leaving the
cores  of the  merging galaxies  on a  bound orbit.   A  rapid merging
follows next, leading to the formation of a single object which is far
from  virial   equilibrium  right   after  the  merging   happens.   A
reorganisation of the structure,  orbital composition of the stars and
shape of  the object  takes place during  some time, leaving  behind a
prolate non-rotating remnant when  the simulation is stopped after the
central parts of  the remnant are close to  virial equilibrium (bottom
row of Fig.~\ref{fig1}).

Model M2  (Fig.~\ref{fig1} middle  column) is an  equal-mass encounter
between two elliptical galaxies, which follow a parabolic orbit with a
large impact parameter. This  simulation differs from the previous one
in  the type  of  encounter.  While  for  model M1  the encounter  was
head-on, model M2 corresponds to  a grazing encounter. Again, prior to
the first pass  through the pericenter there is  not much evolution in
the two  galaxies.  After  the first encounter  and due to  the acting
tidal forces,  the galaxies feature extended plumes  and broad bridges
of particles between  the two bodies. Like in model  M1, a fraction of
the orbital energy is transfered to the particles as kinetic energy. A
number of  particles acquire positive binding energy  and thus escape.
After a second pass through the pericenter the galaxies finally merge.
The orbital angular momentum of the initial galaxies is transferred as
internal angular  momentum to  the stars of  the merged system.   As a
consequence the final  remnant presents a fair amount  of rotation and
due  to the  gentle merging  orbit the  final shape  of the  object is
rather oblate.

These two systems and merging processes give as a result remnants that
match the properties found  in present-day, high luminosity elliptical
galaxies.   Therefore,  this  kind  of dry-merger  between  elliptical
galaxies has  been proposed as a formation  mechanism for present-day,
high luminosity ellipticals~\cite{khochfar05,GG05a}.

Model M3  (Fig.~\ref{fig1} right column) corresponds  to the collision
between two disk-bulge-halo models  with their spins partially aligned
with the orbital angular momentum.  Not much evolution of the disks is
seen  prior  to the  first  encounter,  although  some distortion  and
heating  is  observed.  However,  after  the  first  pass through  the
pericenter (middle row)  prominent tails and a bridge  between the two
central bodies are  created due to the spin-orbit  coupling. The tidal
tails carry away a number of  stars that transport a large fraction of
the angular momentum to the outer parts. After two more passes through
the  pericenter new  tails  are  formed and  finally  the main  bodies
merge. Shells and ripples are still  visible in the outer parts of the
remnant after  the inner  parts have merged  (bottom row).   The large
bulges of the  initial systems help to stabilise  the disks during the
merging stages, as discussed  by~\cite{GG05c}, so right after the last
encounter and the merging a large fraction of the disk particles still
have disk kinematics, leaving a clear signature in the merger remnant.
Both the orbital angular momentum  and the disks spin angular momentum
are transferred to the inner parts of the remnant so the kinematics of
this system  is that  of an  oblate rotator and  the isophotes  of the
remnant present  disky deviations from perfect  ellipses. This process
has  been   proposed  as   a  way  to   build  present-day   mid-  and
low-luminosity elliptical galaxies~\cite{naab03,GG05c}.

\subsection{Gravitational radiation: waveforms and energetics}

\begin{figure}
\includegraphics[width=77mm]{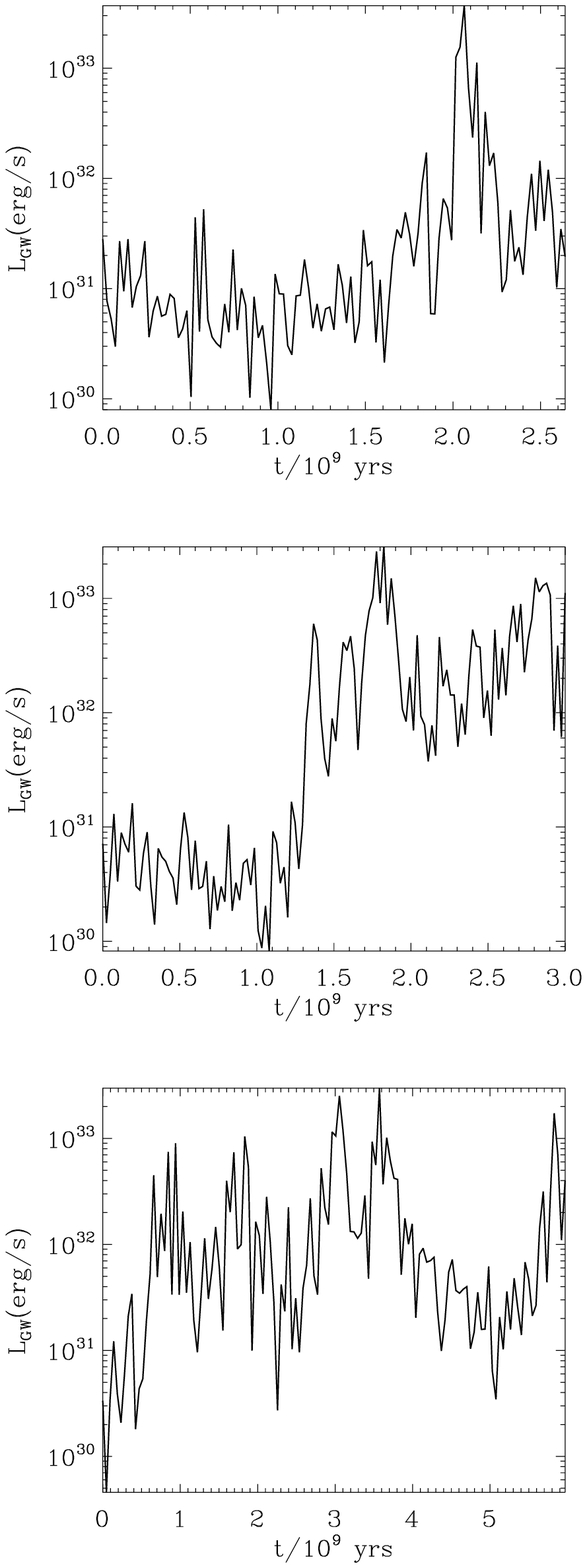}
\caption{Gravitational wave  luminosity as a function of  time for the
three models: M1 (top), M2 (middle), and M3 (bottom).}
\label{fig3}
\end{figure}

We turn now  to describe the gravitational radiation  generated in the
galaxy encounters discussed  in the preceding section. Fig.~\ref{fig2}
shows the dimensionless  GW strain $h$ (only the  ``+" polarisation is
plotted) as a  function of time for the  three models. These waveforms
have been computed using Eq.~(\ref{hij}). This equation shows that the
strain decays  linearly with  the distance.  In  order to  compute the
curves of Fig.~\ref{fig2} we have  assumed a distance of $D=10$ Mpc to
the  source,  which is  compatible  with  observable galactic  systems
undergoing merger  events (e.g.  NGC  4424, The Antennae  galaxies NGC
4038/9 or NGC 520) located  at  few tens  of Mpc. We have checked that
the  amplitude  scales  with  the  distance  as  expected,  while  the
frequency  remains unaffected. The  mergers we  are studying  occur on
characteristic  timescales  of a  few  $10^9$  years.   Note that  the
radiation  signal reaches  its maximum  amplitude in  such timescales.
Therefore, such immense wavelengths lead to tiny GW frequencies of the
order of $\sim 10^{-16}$ Hz.

From the very  beginning of the simulations there  is a non-negligible
contribution to the GW signal associated with the internal dynamics of
the  halos. This  is  due  to the  fact  that the  halos  are made  of
collisionless particle distributions which  are not in equilibrium and
possess a nonzero quadrupole moment. All three waveforms are dominated
by  a maximum  burst-like signal  which is  correlated with  the final
plunge of the two galaxies,  taking place at times $\sim 2\times 10^9$
(top),  $\sim  1.8\times  10^9$  (middle),  and  $\sim  3\times  10^9$
(bottom) years,  for models  M1, M2, and  M3, respectively.  Since the
mergers extend  considerably in time, even though  the final snapshots
in Fig.~\ref{fig1} show virialized  systems (for their central parts),
the  waveforms  keep memory  of  the  violent  events which  leads  to
noticeable variations being still visible  in the GW signal by the end
of the simulations, particularly  for model M1.  This effect, however,
does not have any implication on the typical signal amplitudes ($h\sim
10^{-13}$) and frequencies attained.

The different dynamics of the mergers in models M1 and M2 is reflected
in  the   waveform  patterns  shown   in  the  upper  top   panels  of
Fig.~\ref{fig2}. In both cases  the plunge phase is clearly noticeable
in the burst  signal.  Model M1 corresponds to  a head-on collision, a
process somehow  more violent and  shorter than that present  in model
M2, which describes a non-zero impact parameter encounter leading to a
grazing  collision. Despite  the intrinsically  different evolutionary
tracks  in  models  M1  and   M2,  the  waveform  gross  features  are
similar. This  could be explained  by the complexity of  the scenario,
and  the way  in which  the waveforms  are computed  by adding  the GW
contribution  from each  dark matter  particle,  which may  lead to  a
complex  interference pattern.  Therefore, the  different  dynamics in
both models is hard to disentangle in the GW signal.

On the other hand model M3 corresponds to a disk-like encounter. These
system are inherently more aspherical and, from the early phase of the
evolution, the signal is stronger than in cases M1 and M2. As a direct
consequence of  the morphology of  the disk-like galaxies  involved in
this encounter,  tidal disruptions highly  alter the disks for  a more
extended period of time than in the other two cases. Tidal streams are
visible in the middle panel of Fig.~\ref{fig1} for model M3 (after the
first pass  through the pericenter),  which translates in  a modulated
high-amplitude GW  signal during most  of the simulation.   The signal
peaks at  $t\sim 3.5  \times 10^9$  years which, as  in the  other two
cases, corresponds to the plunge  event. After this, the signal decays
keeping a high amplitude and variability due to the oblate geometry of
the  final object (see  the spiral  structure in  the bottom  panel of
Fig.~\ref{fig1} for model M3).

It  is  worth pointing  out  that  since the  values  of  $h$ and  the
frequency of  the GWs resulting  from our simulated  galaxy encounters
are of  the order  of those expected  from relic GWs  (see e.g.~Fig.~2
of~\cite{baskaran06}), their  effect on  the CMB polarisation  is also
likely to exist.

We  note  that  our results,  in  terms  of  both GW  frequencies  and
amplitudes,  differ  from  those of~\cite{carbone06},  being  slightly
larger for the case of galaxy encounters studied here. The explanation
is to be  found in the fact that~\cite{carbone06}  study the formation
and evolution of isolated, idealised, dark matter halos. No encounters
are considered and  the contribution to the GW  signal comes only from
the quiescence evolution of such halos.

On  the  other hand,  we  have not  included  in  our simulations  the
presence of  SBHs in the numerical  modelling.  This has  been done on
purpose, motivated  by the interest in searching  for the contribution
to the  GW signal  from the collisionless  component of  the galaxies,
leaving aside  the contribution from  the SBHs. This has  no influence
for the signal we are studying  since it is well-known that binary SBH
mergers  show  chirp  waveforms  whose  frequencies  fall  within  the
sensitivity range  of LISA~\cite{holz04}  and, hence, contribute  in a
extremely different band of the GW spectrum.

Figure~\ref{fig3}  shows  the  time  evolution  of  the  gravitational
luminosity of the three models  of our sample. The curves are computed
using  Eq.~(\ref{lum}), reducing  the  order of  the time  derivatives
through  the use  of  the equations  of  motion.  The  first thing  to
mention is the correlation between  the maxima of the luminosities and
the  GW   amplitudes.   As  happens   with  the  signal   strain  (see
Fig.~\ref{fig2}) the luminosity in model M3 shows increased modulation
than for  models M1 and  M2.  We estimate  that the average  GW energy
emitted during the whole duration of the simulations for all models is
$\sim 10^{49}$ erg.

These results show that direct detection of GWs from galaxy mergers is
entirely  inaccessible  to observation.  In  addition,  we note  that,
unfortunately,  gravitational  waves  from  galaxy  encounters  cannot
themselves polarize the CMB in a significant manner. An estimate based
on ~\cite{portilla}, which  is only valid for a  source located in the
wave zone (e.g.~100 Mpc), produces in our case a rotation angle of the
polarization  vector  of the  CMB  photons  $\sim10^{-13}$ radians,  a
completely negligible effect.  On the  other hand, as mentioned in the
introduction  it has been  recently claimed~\cite{carbone06}  that the
distribution of dark matter halos with masses between $5 \times 10^{9}
M_{\odot}$ and $10^{15} M_{\odot}$  produces a background of GWs whose
effects  on  the  CMB  polarization  deserve  a  detailed  study.   In
particular, B-mode  polarization could be detected  in future missions
tailored  to study  this  polarization mode.   This  issue deserves  a
detailed study which will be presented elsewhere.

\section{Summary}
\label{summary}

In  this paper  we have  studied the  gravitational waves  produced by
galaxy encounters, paying attention  to the collisionless component of
the systems.  To  the best of our knowledge  such an investigation has
not been carried out before. For our study we have performed numerical
simulations of a sample of representative galaxy encounters, for which
we  have  extracted,  using  the  Newtonian  quadrupole  formula,  the
gravitational waveforms,  amplitudes and  frequencies, as well  as the
luminosities. The  main result of our  work has been to  find out that
GWs produced in galaxy encounters have typical values of the GW strain
and frequencies which overlap to those arising in strong variations of
the gravitational field in the early Universe. Our simulations confirm
that the absence of the central  SBH does not affect the GW spectra we
have computed,  since the  signals from binary  SBH mergers  have very
different features, in  terms of frequency and amplitude,  to the ones
presented in this paper.

Recently, a number of missions have been proposed to detect signatures
of  GWs (with  wavelengths comparable  to  the size  of the  Universe)
produced  by quantum  fluctuations of  spacetime during  inflation, by
measuring the weak imprint they  leave on the polarization of the CMB.
Given the  fact that the main features of our results could fit in the
expected range  of detectability of those missions,  we have estimated
the effects of a close galaxy encounter on the CMB polarization. Thus,
we have investigated this possibility  to find out whether this effect
could add  an unexpected  source of GW  ``noise" which may  hinder the
original target  of the experiments, namely  relic gravitational waves
from inflation.

The  most appealing  motivation of  this  work could  be the  possible
direct detection  of dark  matter in galaxy  halos.  By focusing  on a
particular galaxy  encounter if the  polarization of CMB  photons were
modified, this would imply that  the source responsible of such effect
would  be  the expected  dark  matter  distribution.  Therefore,  dark
matter  should exist  in concordance  with the  so-called  dark matter
paradigm.  The  detection of those elusive GWs  (through their imprint
on the CMB polarization) would  provide a smoking-gun signature of the
existence of  the dark matter.  However, the results of  our estimates
show that the gravitational-wave-induced imprints of galaxy encounters
on the CMB are beyond observational capabilities.  Nevertheless, there
exist additional  aspects of the nonlinear dynamics  of galaxy mergers
which  could  have  implications  on  potentially  observable  effects
through CMB polarization, namely gravitational lensing~\cite{Hu-White}
or  the so-called Stroskii  effect (see~\cite{Morales}  and references
therein). This issue will be discussed in a forthcoming paper.

The results reported  in this work are similar  to those investigated
by~\cite{quilis98}   and~\cite{carbone06}   regarding  the   nonlinear
evolution  of  cosmological  structures  as sources  of  gravitational
radiation. Future work on this topic should account for the integrated
effect of  a realistic  galaxy distribution. The  hopes opened  by the
upcoming  missions  to  measure  ultra-low frequency  GWs  make  these
investigations worth to pursue.

\begin{acknowledgements}
The authors thank Alicia Sintes, Miguel Portilla, Juan A.~Morales, and
Jos\'e Mar\'{\i}a Ib\'a\~nez for useful discussions, and the anonymous
referee for the stimulating insights and comments.  Research supported
by the Spanish {\it Ministerio  de Educaci\'on y Ciencia} (MEC; grants
AYA2006-02570,  AYA2006-08067-C03-01,  and  FIS2006-06062).  VQ  is  a
Ram\'on y Cajal Fellow of the Spanish MEC.
\end{acknowledgements}

\end{document}